\documentclass[
    ,final            % use final for the camera ready runs
  ]
  {aipproc}

\layoutstyle{8x11double}

\begin{document}

\title{Finding pulsars with LOFAR}

\classification{97.60.Gb}
\keywords      {pulsars: general --- telescopes --- surveys}

\author{Joeri van Leeuwen}{
  address={ Department of Astronomy, University of California, 601 Campbell
  Hall, Berkeley CA 94720, USA \footnote{joeri@astro.berkeley.edu} }
}

\author{Ben Stappers}{
  address={ Stichting ASTRON, PO Box 2, 7990 AA Dwingeloo, The
  Netherlands }
  ,altaddress={ University of Manchester, Jodrell Bank Observatory,
  Macclesfield, Cheshire SK11 9DL, UK }
}

\begin{abstract}
We investigate the number and type of pulsars that will be
discovered with the low-frequency radio telescope LOFAR. We consider
different search strategies for the Galaxy, for globular clusters and
for galaxies other than our own. We show an all-sky Galactic survey
can be optimally carried out by {\em incoherently} combining the 
LOFAR stations. In a 60-day all-sky Galactic survey LOFAR can 
find over a thousand
pulsars, probing the local pulsar population to a very deep
luminosity limit. For targets of smaller angular size, globular
clusters and galaxies, the LOFAR stations can be combined coherently,
making use of the full sensitivity. Searches of nearby northern-sky globular
clusters can find large numbers of low luminosity millisecond pulsars
(eg.\ over 10 new millisecond pulsars in a 10-hour observation of
M15). If the pulsar population in nearby galaxies is similar to that
of the Milky Way, a 10-hour observation can find the 10 brightest
pulsars in M33, or pulsars in other galaxies out to a distance of
1.2Mpc. 
\end{abstract}

%%%%%%%%%%%%%%%%%%%%%%%%%%%%%%%%%%%%%%%%%%%%%%%%%%%%%%%%%%%%%%%%%%%
%%
%% The below \maketitle command inserts the actual front matter data.
%% It has to follow the above declarations.
%%
%%%%%%%%%%%%%%%%%%%%%%%%%%%

\maketitle

%%%%%%%%%%%%%%%%%%%%%%%%%%%%%%%%%%%%%%%%%%%%
%% MAINMATTER
%%
%%%%%%%%%%%%%%%%%%%%%%%%%%%%%%%%%%%%%%%%%%%%%%%%%%%%%%%%%%%%%%%%%%%%%%%%%%%%
%% Headings:
%%
%% The aipproc supports three heading levels, i.e., \section,
%%	\subsection, and \subsubsection.
%%%%%%%%%%%%%%%%%%%%%%%%%%%%%%%%%%%%%%%%%%%%%%%%%%%%%%%%%%%%%%%%%%%%%%%%%%%%
%% Cross-references:
%%
%% Page numbers (\pageref) and headings can NOT be referenced in the class,
%% since before being produced, no page numbers are determined.
%%
%% Tables, figures, and equeations can be referenced by using the LaTex
%% 	commands \label and \ref. For references to equation numbers, \eqref
%%	can be used, which will print "(1)" (while \ref will result in "1").
%%
%%%%%%%%%%%%%%%%%%%%%%%%%%%%%%%%%%%%%%%%%%%%%%%%%%%%%%%%%%%%%%%%%%%%%%%%%%%%
%% Lists: 
%%
%% Standard "itemize", "enumerate", etc. list environments are supported.
%%%%%%%%%%%%%%%%%%%%%%%%%%%%%%%%%%%%%%%%%%%%%%%%%%%%%%%%%%%%%%%%%%%%%%%%%%%%
%% Urls:
%%
%% \url{} command is provided for documenting URLs.
%%%%%%%%%%%%%%%%%%%%%%%%%%%%%%%%%%%%%%%%%%%%

\section{Introduction}

Since the discovery of the first four pulsars with the Cambridge radio
telescope, an ongoing evolution of telescope systems
has doubled the number of known radio pulsars roughly every 4 years.

The next step in radio telescope evolution will be the use of large
numbers of low-cost receivers that are combined to form an
interferometer or a single dish. These telescopes, LOFAR \citep{ls07},
the Allen Telescope
Array \citep{bow07} and the SKA \citep{kram04}, create new
possibilities for pulsar research.

Here we outline and compare strategies for
targeting normal and millisecond pulsars, both in the disk and
globular clusters of our Galaxy, and in other galaxies.

%----------------------------------------------------
\section{LOFAR - The Low Frequency Array}
\label{sec:LOFAR}
%% how does the LOFAR idea work -- phased array

With the first test station operational and the first pulsar detected,
LOFAR is on track to start operation in 2008. We have evaluated and
simulated the LOFAR reference configuration for pulsar searches, and will
describe the outcome here. For details, see \citet{ls07}.

Using two different types of dipoles, LOFAR can observe in both a low
and a high band, ranging from 30-80\,MHz and 110-240\,MHz
respectively. Because of its higher sensitivity, we only discuss using
the high-band antennas (HBAs) here.

The basic collecting elements are the individual dual-polarization
dipoles; a 4x4 tile of these is defined as 1 antenna.  Sets of 64
antennae are grouped together in stations that are $\sim$60\,m
across. For the compact core, which encompasses the 24 stations in the
inner 2\,km of the telescope, the {\em Stella} supercomputer forms 8
to 32 independently steerable beams on the sky, each with a bandwidth
of 32\,MHz.

For the high-band antennas the effective area towards zenith is
28\,m$^2$ per antenna at 120\,MHz (for a total theorical gain of about
16K/Jy), halving at a zenith angle of 60$^\circ$. Toward higher
frequencies both quantities decrease until at 240\,MHz the maximum
effective area is 8\,m$^2$ per antenna.

\begin{figure}
  \includegraphics[width=0.48\textwidth]{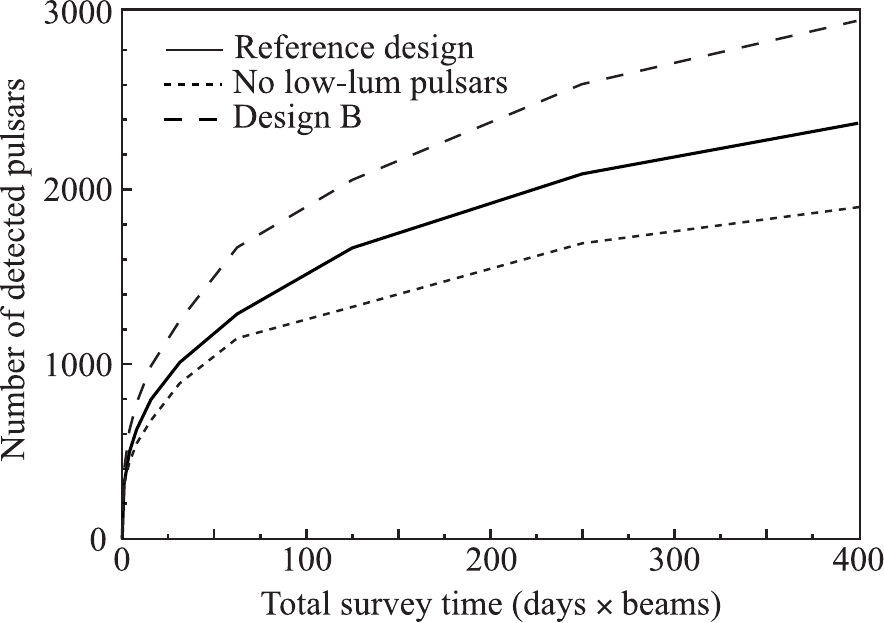}
  \caption{
	   Simulated number of pulsars detected in an all-sky survey, versus total survey time
   in days $\times$ beams. About $\frac{3}{4}$ of these detections are new discoveries, a ratio that is
   relatively independent of integration time or observing frequency.
	}	
  \label{img:n-vs-t}
\end{figure}%

%----------------------------------------------------
\section{Pulsar searches}
\subsection*{Galactic disk surveys}

For 'sparse' telescopes incoherent adition of the stations
allows for longer integration times, to first order making the
LOFAR incoherent case about 6 times more sensitive than the coherent
approach for the same total amount of time spent \citep[cf.][]{bac99}.

Although the sky background and scatter broadening increase towards
lower frequency, the total sensitivity is mainly determined by the
effective collecting area and therefore 120MHz is the most efficient
frequency for the survey.

For our reference-design simulation we conservatively use a gain of
0.66 times the theoretical value and although the testing has shown
the radio-interference environment to be relatively clear, we shall
here only assume 80\% of the 32MHz band to be free of RFI (for
details, see  \citet{ls07}. Design B uses a perhaps more realistic
efficiency of 0.75.

We use a
population synthesis code that simulates the birth, evolution, death
and possible detection of radio pulsars, as described more extensively
in \citet{bwhv92, lv04}.

\begin{figure}[t]
   \includegraphics[width=\textwidth]{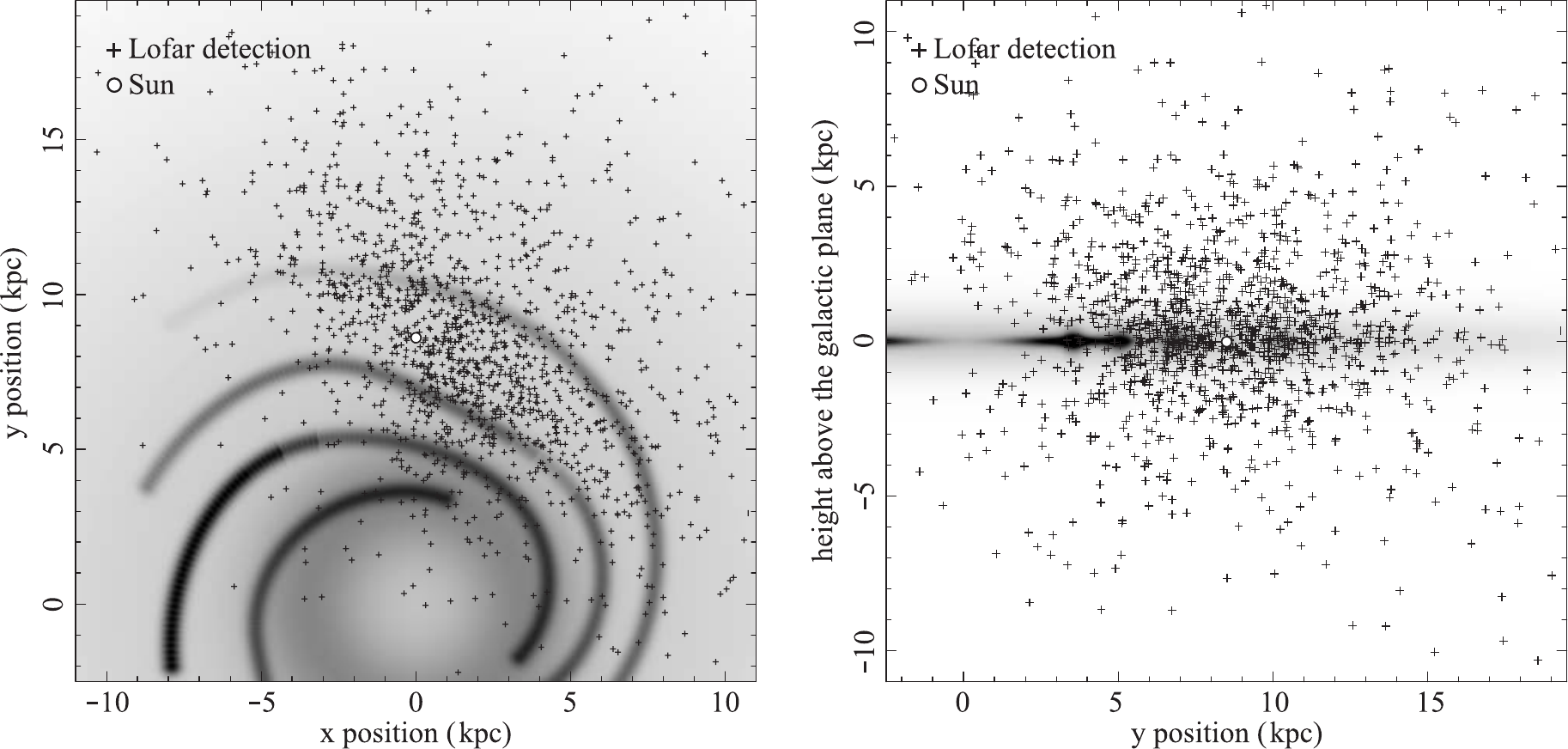}
   \caption{
	The 1000+ pulsars discovered in a 60-day LOFAR all-sky survey simulation.
  	ISM shown in gray. Left) projected on the
  	Galactic plane. Right) projected on the plane through the
	Galactic centre and sun, perpendicular to the disk.
	   }
  \label{img:x-y-z}
\end{figure}

For 8 pulsar surveys (Jodrell,  UMass-Arecibo, MolongloII,
UMass-NRAO, ParkesII, Cambridge 80MHz, Parkes Multibeam and LOFAR) we model
the sensitivity function versus period, dispersion and scattering
smearing, pulse width, and sky position.  By comparing the simulated
pulsar sample with the real one for the first four surveys
\citet{hbwv97a} determined the most probable underlying initial model
parameters. We use this best model to determine the yield of a future
surveys with LOFAR. As a test, we also model the second 81.5-MHz
Cambridge survey \citep{stw98} to check the validity of our
extrapolations to lower frequencies: With 27 $\pm$ 5 simulated pulsars
found, our model reproduces their actual tally of 20 pulsars quite well.

For LOFAR, we have evaluated surveys for different pointing duration
and different telescope parameters. In our reference design
LOFAR can find over a thousand pulsars in a 60-day survey on a single beam.

Compared to the Parkes Multibeam survey, the pulsars found with LOFAR
are different in several ways. Because of its higher sensitivity,
LOFAR detects more low-luminosity nearby pulsars. If the 1mJy kpc$^2$
lower limit at 400MHz exists, LOFAR can even see all pulsars that are
beamed towards us, up to 1.4 kpc. If it does not, then this survey
will certainly oncover the actual low-luminosity distribution: a
number of importance if one wants to understand the neutron-star
birthrate. The same goes for even the fastest relatively nearby millisecond pulsars.

\subsection*{Galactic globular cluster surveys}

In a Galactic disk survey, one trades field of view for sensitivity by
adding the signal of the stations incoherently instead of
coherently. Specific smaller regions on the sky with higher densities
of radio pulsars can be targeted with smaller field of view,
but significantly higher sensitivity.

Globular clusters fit the description well; they are compact and form
regions on the sky with high stellar densities. These high
densities also cause globular clusters to contain more binaries and
binary-products than are found in the disk. This makes globular
clusters very good candidates for millisecond pulsar
searches.

At 120MHz, MSPs with periods of around 2--5 ms
are detectable up to a DM of about 50 pc/cm$^3$, after which
scattering becomes a problem. Observing at
200MHz extends this limit to 80 pc/cm$^3$. Promising cluster M15 is very compact and fits
well into a coherent compact-core beam\citep{ls07}. We
can now use the full gain of the coherent addition, which makes LOFAR
at least twice as sensitive as the Arecibo survey that found the other
milliseconds pulsars in M15 \citep{and92}. If the $\frac{d log L}{d
log N}=-$1 relation extends to lower luminosities, a 10-hour
observation can yield over 10 new millisecond
pulsars. 

\subsection*{Pulsars in other galaxies}

The next obviously dense regions on the sky are galaxies. Compared to
globular clusters, galaxies have several advantages for a LOFAR pulsar
survey. They are well represented in the northern hemisphere and if
visible face-on and located in the part of the sky that is pointed
away from our Galactic disk, the scatter broadening is relatively low.
In a relatively close galaxy like M33, a 10 hour pointing with the
compact core at 140MHz could detect all pulsars more luminous than
57Jy kpc$^2$. In our own Galaxy, ten pulsars
of such luminosity are known. We could observe the brightest pulsars
in our Galaxy (like B1302$-$64, 130Jy kpc$^2$) to distances of about
1.2Mpc.

%----------------------------------------------------
\section{Discussion}

\subsection*{Steep spectrum sources}
The generally steep spectrum of pulsars typically turns over at lower
frequency, between 100 and 250MHz \citep{mgj+94}. There is, however, a
small fraction of pulsars for which no such spectral break has been
observed, to frequencies as low as 50MHz. Sources with a spectral
index steeper than the spectral index of the sky background of $-$2.6
\citep[e.g.\ PSR~B0943+10, spectral index $-$4.0][]{rd94} will be more
easily detectable by a telescope like LOFAR, producing
quantitative input for radio emission models.

\section{Conclusions}
Because of its large area and wide beam on the sky, LOFAR probes the
local population of pulsars to a very deep luminosity limit. A 60-day
Galactic survey at 140MHz can find over a thousand new pulsars,
disclosing the local low-luminosity population. If we add all antennae
coherently the sensitivity increases even further; with this setup,
millisecond pulsars in nearby globular clusters can be detected to
much lower flux limits than previously possible. Assuming the pulsar
population in other galaxies is similar to that in ours, we can detect
periodicities or giant pulses from extragalactic pulsars up to several
Mpcs away.

\bibliographystyle{aipproc}   % if natbib is available


\bibliography{journals_apj,modrefs,psrrefs,crossrefs,modjoeri,psrfrank}
%\bibliography{sample}

%%%%%%%%%%%%%%%%%%%%%%%%%%%%%%%%%%%%%%%%%%%%%%%%%
%% If the bibliography is
%% produced without BibTeX, comment out the above lines, use
%% \begin{thebibliography}{widest-label} environment to hold 
%% the list of references and 
%% \bibitem{label} command to start a bibliographical entry having
%% the "label" for use in \cite commands.

\end{document}